%% file: main.tex
\documentclass[10pt,twocolumn,letterpaper]{article}

\usepackage{iccv}
\usepackage{times}
\usepackage{epsfig}
\usepackage{graphicx}
\usepackage{amsmath}
\usepackage{amssymb}
\usepackage{algorithm}
\usepackage{algorithmic}
\usepackage{bbm}
\usepackage{amsfonts}
\usepackage{mathrsfs}
\usepackage{multirow}

\usepackage[pagebackref=true,breaklinks=true,letterpaper=true,colorlinks,bookmarks=false]{hyperref}

\iccvfinalcopy 

\def\iccvPaperID{9} 
\vspace{-20pt}
\ificcvfinal\fi

\begin{document}

\title{Unimodal-uniform Constrained Wasserstein Training for Medical Diagnosis}

\author{Xiaofeng Liu{$^{1,2}$},~ Xu Han{$^{3\dag}$},~ Yukai Qiao{$^{4\dag}$},~ Yi Ge{$^{2\dag}$},~ Site Li{$^{2}$},~ Jun Lu{$^{1*}$}\\\vspace{-8pt}{\small~}\\
{$^{1}$}BIDMC, Harvard Medical School;~~~ {$^{2}$}Carnegie Mellon University;\\ {$^{3}$}Johns Hopkins University;~~~ {$^{4}$}Beijing University of Posts and Telecommunications\\
{\small{{$^{\dag}$}Contribute equally~~{$^{*}$}Corresponding Author: \tt{{jlu@bidmc.harvard.edu}}}}
}

\maketitle

\input{1_Abstract.tex}

\input{2_Introduction.tex}

\input{3_RelatedWork.tex}

\input{4_Approach.tex}

\input{4_Approach1.tex}

\input{4_Approach2.tex}

\input{4_Approach3.tex}

\input{5_Experiments.tex}

\input{6_Conclusions.tex}

{\small
\bibliographystyle{ieee}
\bibliography{egbib}
}

\end{document}

%% file: 1_Abstract.tex
\begin{abstract}

The labels in medical diagnosis task are usually discrete and successively distributed. For example, the Diabetic Retinopathy Diagnosis (DR) involves five health risk levels: no DR (0), mild DR (1), moderate DR (2), severe DR (3) and proliferative DR (4). This labeling system is common for medical disease. Previous methods usually construct a multi-binary-classification task or propose some re-parameter schemes in the output unit. In this paper, we target on this task from the perspective of loss function. More specifically, the Wasserstein distance is utilized as an alternative, explicitly incorporating the inter-class correlations by pre-defining its ground metric. Then, the ground metric which serves as a linear, convex or concave increasing function w.r.t. the Euclidean distance in a line is explored from an optimization perspective.

Meanwhile, this paper also proposes of constructing the smoothed target labels that model the inlier and outlier noises by using a unimodal-uniform mixture distribution. Different from the one-hot setting, the smoothed label endues the computation of Wasserstein distance with more challenging features. With either one-hot or smoothed target label, this paper systematically concludes the practical closed-form solution. We evaluate our method on several medical diagnosis tasks (e.g., Diabetic Retinopathy and Ultrasound Breast dataset) and achieve state-of-the-art performance.

\end{abstract}

%% file: 2_Introduction.tex
\section{Introduction}

In the realm of medical diagnosis, there are numerous prediction tasks in which the output labels demonstrate high discrete and successive features. The problem of the health risk level could be a very example. Although it can be a continuous variable, it is often discretized $i.e.$, at several intervals in practices. In parallel with previous statements, the Diabetic Retinopathy Diagnosis (DR) labels the disease level into five rankings: no DR (0), mild DR (1), moderate DR (2), severe DR (3) and proliferative DR (4) \cite{liu2018ordinal,liu2019unimodala}. This kind of labeling system has been widely accepted by The Breast Imaging-Reporting and Data System (BIRADS), liver (LIRADS), gynecology (GIRADS), colonography (CRADS), etc. 

Due to the successively nature of risk level, the error that misclassifying a proliferative DR (4) image to mild DR (1) is considerably severe than the counterpart that misclassifying to severe DR (3). According to earlier literature, this kind of medical diagnosis task often casts as a multi-class classification problem or a metric regression problem.

\begin{figure}[t]
\centering
\includegraphics[width=6.5cm]{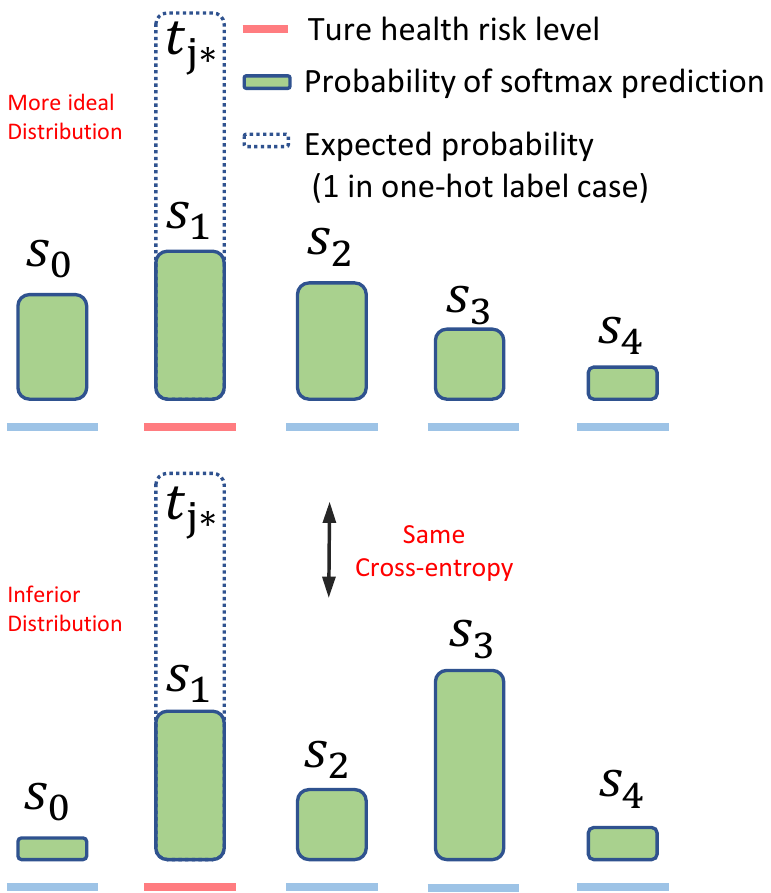}\\
\caption{The limitation of CE loss for health risk level estimation. The ground truth class is $t_j \ast$. Two possible softmax predictions (green bar) of the health risk level estimator have the same probability at $t_j \ast$ position. Therefore, both predicted distributions have the same CE loss. However, the top prediction is preferable to the bottom, since we desire the predicted probability distribution to be larger and closer to the ground truth class.}\label{fig:1} 
\end{figure}

The multi-class classification formulation using the cross-entropy (CE) loss, the class labels are assumed to be independent of each other \cite{liu2019conservative}. Therefore, the inter-class similarity is not properly exploited. For instance, in Fig. \ref{fig:1}, the histogram prediction is most well-liked to be focused close to the truth category, whereas the Cross entropy loss doesn't encourage that.

On the opposite hand, regression treats the discrete risk as continuous value. As discussed in \cite{liu2019conservative,niu2016ordinal,mahendran2018mixed,divon2018viewpoint}, training the regression model with discrete labels can result in over-fitting and has similar or lower accuracy than classification. Therefore, it's necessary to contemplate the sequential and discrete nature of the ordinal medical diagnosis.

Recent works generally use the $N-1$ binary classification sub-tasks using sigmoid output with MSE loss or softmax output with cross entropy loss, where $N$ is the number of levels. Unfortunately, the cumulative probabilities $p(y>1|\textbf{\texttt{x}}),...,p(y>N-1|\textbf{\texttt{x}})$ are calculated by several independent branches, failing to guarantee the that they are monotonically decreasing. This result in $p(y=i|\textbf{\texttt{x}})$ don't seem to be ensured to be strictly positive, additionally the poor learning potency within the early stage of training. Moreover, $N-1$ weights need to be manually fine-tuned to balance the CE loss of each branch. \cite{liu2018ordinal,liu2019unimodala} proposes to use a stick-breaking process to re-parametrize the outputs of $N-1$ units that is associated with the Bayesian non-parametric Dirichlet process. Embarking on this, the cumulative probabilities can achieve the expected monotonical decrease, but it couldn't be unheeded that it's considerably more sophisticate than standard CE-loss.

Furthermore, \cite{da2008unimodal} propose to use one output unit to calculate the parameter of a unimodal distribution, and define the $p(y=i|\textbf{\texttt{x}})$ follows a Poisson or Binomial distribution, which suffers from lacking the ability to control the variance. Since the peak (also the mean and variance) of a Poisson distribution is equal to a designated $\lambda$, the peak cannot be assigned to the first or last class, and its variance is intended to be rather high when the peak is needed in the very later classes.

In this paper, we employ the Wasserstein loss as an alternative for empirical risk minimization. The $1^{st}$ Wasserstein distance is defined as the cost of optimal transport for moving the mass in one distribution to match the target distribution \cite{bogachev2012monge,rubner2000earth,ruschendorf1985wasserstein}. Specifically, the Wasserstein distance, between softmax prediction and its target label that are separately  normalized as histograms, is measured. By defining the ground metric as class similarity, prediction performance earns the measure room in a sensitive way to correlations between the classes.

The ground metric can be predefined when the similarity structure is known as a priori to incorporate the inter-class correlation, $e.g.,$ the Euclidean distance in a line. We further extend the Euclidean distance to its increasing function from an optimization perspective \cite{liu2018joint,liu2017line}. The exact Wasserstein distance in a one-hot target label setting can be formulated as a soft-attention scheme of all prediction probabilities and be rapidly computed.

Another challenge of health risk level estimation stems from the label quality. For instance, the agreement rate of the radiologists for malignancy is usually less than 80\%, resulting in a noisy labeled dataset \cite{nishikawa2016agreement,salazar2017reliability}. Despite the often-uncleared distinction between adjacent labels, it is more possible that a well-trained annotator will mislabel a Severe DR (3) sample to Moderate DR (2) rather than No DR (0). This requires modeling the noise for robust training \cite{huber2011robust,che2019deep}.

The mis-labeled data can misleading the training \cite{szegedy2016rethinking,bekker2016training,belagiannis2015robust,pereyra2017regularizing}. In here, two kinds of the label noise are investigated. The outlier noise refer to a sample is different from all of the others which is caused by the random error. It can be sculpturesque by a uniform distribution \cite{szegedy2016rethinking}. However, it can be more common that the ordinal case is a lot of probably to possess the inlier noise, in which the labels are wrong annotated as the near one. We propose to to model it with a unimodal distribution. We give a solution that use a smoothed target histogram by smoothing the one-hot label with an uniform-unimodal mixture model.

In contrst to the one-hot setting, the smoothed label makes the calculation of Wasserstein loss more sophisticated. This is  due to the various attainable transportation plans. The $\mathcal{O}(N^3)$ computational complexity for $N$ classes has long been difficulty in using Wasserstein distance for large-scale applications. Rather than approximate its Wasserstein distance with a $\mathcal{O}(N^2)$ complexity algorithm \cite{cuturi2013sinkhorn,frogner2015learning}, we propose to systematically conclude the fast closed-form computation of Wasserstein distance in the setting of smoothed label, when the ground metric is a linear, convex, or concave increasing function $w.r.t.$ the Euclidean distance. We show that the linear and convex cases can be solved with a linear complexity of $\mathcal{O}(N)$. In comparison to its approximate counterpart, the $exact$ solutions that this paper proposes are more effective.

The main contributions of this paper are summarized as follows:

$\bullet$ The health risk level estimation casts as a Wasserstein training problem. The inter-class relationship of health risk level data is explicitly incorporated as prior information in the gained ground metric which can be pre-defined ($e.g.,$ a function $w.r.t.$ Euclidean distance in a line).

$\bullet$ The inlier and outlier error of health risk level is modeled with a discrete unimodal-uniform mixture distribution, and regularizes the target confidence by transforming one-hot label to the smoothed target label.

$\bullet$ For either one-hot or smoothed target label, this paper systematically concludes the possible fast closed-form solution when a non-negative linear, convex or concave increasing mapping function is applied in ground metric.  

We validate the the proposed method on several medical diagnosis datasets. Our method acheieves the state-of-the-art performance over the current methods, especially choosing the convex mapping function for ground metric, smoothed target, and closed-form solution.

%% file: 3_RelatedWork.tex
\section{Related Works}

\subsection{Health risk level estimation}

The traditional ordinal level estimation methods can be classified to 3 categories, $i.e., $ naive, binary decomposition and threshold methods \cite{gutierrez2016ordinal,ma2019convex,zhao2015ordinal}. Moreover, the health risk isn't the sole area that has the successive and discrete label. The age prediction and atheistic rating also tightly related to this task. Following the development of deep learning, several works have been proposed to target the successive data. \cite{liu2018ordinal,ratner2018learning} put forward the multi-task learning framework. However, the percentages of each class are not guaranteed to be positive, which may hurt the training, especially
that in the early stage. Besides, there are $N-1$ weights to balance the branches, which is a hard task for manually tuning. \cite{liu2018ordinal,liu2019unimodala} proposes a sophisticated stick-breaking process to reparameterise the $N-1$ outputs to alleviate this issue. \cite{liu2017deep} incorporate the metric learning for data relationship analysis. Different from these methods, we propose to use the Wasserstein distance as the optimization objective to inherit the label similarity.

\subsection{Wasserstein distance}

{Wasserstein distance} is a distance function defined between probability distributions on a given metric space \cite{liu2019conservative}. In these few years, it attracts a lot attention in adversarial generative models $etc$ \cite{arjovsky2017wasserstein,liu2019feature,liu2018data,liu2018normalized,liu2019hard}. It also been used for multi-class multi-label task with a linear model \cite{frogner2015learning}. Since the significant amount of computation costrequired to solve its exact solution, these methods usually choose the approximate solution, of which the complexity is still in $\mathcal{O}(N^2)$ \cite{cuturi2013sinkhorn}. The fast computing of discrete Wasserstein distance is also closely related to SIFT \cite{pele2008linear,cha2002measuring} descriptor, hue in HSV or LCH space \cite{cha2002fast,delon2010fast}. Based on these works, we further adapt this idea to the heath risk level estimation, and encode the correlation of label classes using the ground matrix \cite{liu2019conservative}. We show that the fast algorithms exist in our health risk label structurewith the one-hot or smoothed target label. Moreover, the ground metric used in here does not limited to the Euclidean distance.

\subsection{Robust training with noise data}

{Robust training with noise data} has been studied for a long time in general classification area \cite{huber2011robust,bekker2016training}. The possible solution can be smoothing the one-hot label \cite{szegedy2016rethinking} with a uniform distribution or regularizing the entropy of softmax output \cite{pereyra2017regularizing}. However, for the discrete successive label, the studies speak little voice.

\subsection{Unimodality of Discrete and Successive Data}

\cite{da2008unimodal} propose to predict a Poisson distribution. In their parametric version, the output of the neural network is a single sigmoid unit, which predict the parameter $\lambda$ in a Poisson distribution. However, requiring the output strictly follows a specific distribution could be a very strong assumption. Besides, it is difficult to control the variance of the resulting Poisson distribution. \cite{beckham2017unimodal} introduces an additional temperature parameter to control the variance, but results in more complicate hyper-parameter tuning. Here, we propose to use an exponential function following the softmax to flexibly adjust the shape of target label distribution. Noticing this modification works on the target label instead of the output distribution as \cite{beckham2017unimodal}.

%% file: 4_Approach.tex
\section{Methodology}

\begin{figure}[t]
\centering
\includegraphics[height=5.5cm]{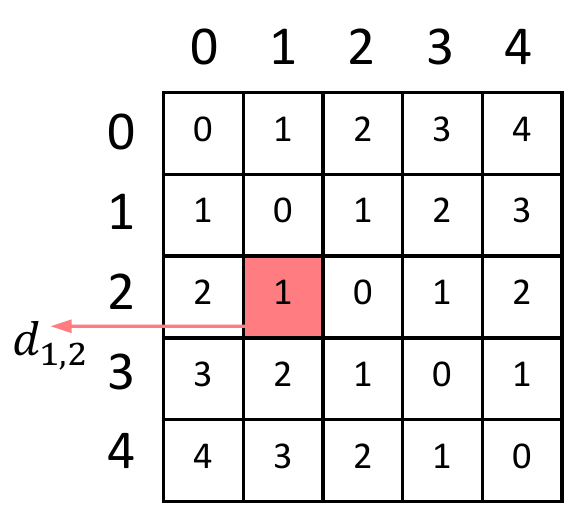}
\caption{The ground matrix using Euclidean distance as ground metric.}
\label{fig:r2}
\end{figure}

In this paper, we consider the task of learning a health risk level estimator ${h}_\theta$, parameterized by $\theta$, with $N$-dimensional softmax output unit. It maps a medical image {\rm\textbf{x}} to a vector ${\rm\textbf{s}}\in\mathbb{R}^N$. We perform learning over a hypothesis space $\mathcal{H}$ of ${h}_\theta$. With the input {\rm\textbf{x}} and its target ground truth one-hot label ${\rm\textbf{t}}$, typically, learning is performed via empirical risk minimization to solve $\mathop{}_{{h}_\theta\in\mathcal{H}}^{\rm min}\mathcal{L}({h}_\theta({\rm\textbf{x}}),{\rm\textbf{t}})$, with a loss $\mathcal{L}(\cdot,\cdot)$ acting as a surrogate of performance measure.

However, cross entropy-based loss treat the output dimensions independently \cite{frogner2015learning}. Therefore, it ignores the similarity structure on label space.  

${\rm\textbf{s}}=\left\{s_i\right\}_{i=0}^{N-1}$ is the output of ${h}_\theta({\rm\textbf{x}})$, $i.e.,$ softmax prediction with $N$ classes (health risk levels). ${\rm\textbf{t}}=\left\{t_j\right\}_{j=0}^{N-1}$ is the target label distribution, where $i,j\in\left\{0,\cdots,{\small N-1}\right\}$ be the index of dimension (class). Assume that the class label possesses a ground metric ${\rm\textbf{D}}_{i,j}$, which measures the similarity of $i$-th and $j$-th dimensions of the output. There are $N^2$ possible ${\rm\textbf{D}}_{i,j}$ in a $N$ class dataset and form a ground distance matrix $\textbf{D}\in\mathbb{R}^{N\times N}$. When ${\rm\textbf{s}}$ and ${\rm\textbf{t}}$ are both histograms, the discrete measure of exact Wasserstein loss is defined as 

\begin{equation}
\mathcal{L}_{\textbf{D}_{i,j}}({\rm{\textbf{s},\textbf{t}}})=\mathop{}_{\textbf{W}}^{{\rm inf}}\sum_{j=0}^{N-1}\sum_{i=0}^{N-1}\textbf{D}_{i,j}\textbf{W}_{i,j} \label{con:df}
\end{equation} where \textbf{W} is the transportation matrix with \textbf{W}$_{i,j}$ indicating the mass moved from the $i^{th}$ point in source distribution to the $j^{th}$ target position. A valid transportation matrix \textbf{W} satisfies: $\textbf{W}_{i,j}\geq 0$; $\sum_{j=0}^{N-1}\textbf{W}_{i,j}\leq s_i$; $\sum_{i=0}^{N-1}\textbf{W}_{i,j}\leq t_j$; $\sum_{j=0}^{N-1}\sum_{i=0}^{N-1}\textbf{W}_{i,j}={\rm min}(\sum_{i=0}^{N-1}s_i,\sum_{j=0}^{N-1}t_j)$.

The entries of ${\rm\textbf{D}}$ in Wasserstein distance are usually unknown, but they have clear meanings in our task. The $i,j$-th entry ${\rm\textbf{D}}_{i,j}$ indicates the geometrical distance between the $i$-th and $j$-th points in a line. A possible choice is using the Euclidean distance ${d_{i,j}}$ of a line ($i.e., \ell_1$ distance between the $i$-th and $j$-th points in a line) as the ground metric $\textbf{D}_{i,j}={d_{i,j}}$.

\begin{equation}
d_{i,j}=|i-j| \label{con:d}
\end{equation}

The Wasserstein distance can be the Earth mover's distance when the two distributions have the same total masses ($i.e., \sum_{i=0}^{N-1}s_i=\sum_{j=0}^{N-1}t_j$) and using the symmetric distance $d_{i,j}$ as ${\rm\textbf{D}}_{i,j}$. The ground matrix using Euclidean distance is shown in Fig.\ref{fig:r2}.

The previous efficient algorithms to solve Wasserstein distance usually holds only for $\textbf{D}_{i,j}={d_{i,j}}$ \cite{cha2002measuring,pele2008linear,rubner2000earth,cabrelli1995kantorovich,werman1986bipartite} and do not consider the neural network optimization. Regarding this, this paper proposes extending the ground metric in ${\rm\textbf{D}}_{i,j}$ as $f(d_{i,j})$, where $f$ is a positive increasing function $w.r.t.$ $d_{i,j}$.

%% file: 4_Approach1.tex
\begin{figure}[t]
\centering
\begin{tabular}{cc}
\includegraphics[height=4.8cm]{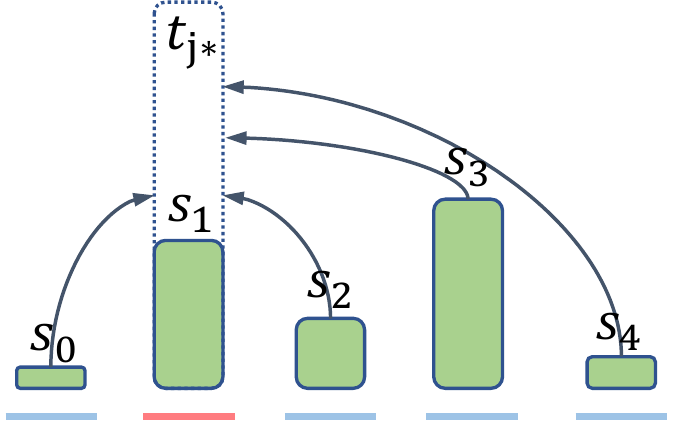}
\end{tabular}
\caption{The only possible transport plan in one-hot target case.}
\label{fig:r3}
\end{figure}

\subsection{Wasserstein training with one-hot target}
In the multi-class single-label dataset, the one-hot encoding is typically used. The distribution of a target label probability is ${\rm\textbf{t}}=\delta_{j,j^*}$, where $j^*$ is the ground truth class, $\delta_{j,j^*}$ is a Dirac delta, which equals to 1 for $j=j^*$\footnote{\noindent We use $i,j$ interlaced for ${\rm \textbf{s}}$ and ${\rm \textbf{t}}$, since they index the same group of positions in a line.}, and $0$ otherwise.

\noindent\textbf{Theorem 1.} \textit{Assume that} $\sum_{j=0}^{N-1}t_j=\sum_{i=0}^{N-1}s_i$, \textit{and} ${\rm{\textbf{t}}}$ \textit{is a one-hot distribution with} $t_{j^*}=1 ($or $\sum_{i=0}^{N-1}s_i)$\footnote{We note that softmax cannot strictly guarantee the sum of its outputs to be 1 considering the rounding operation. However, the difference of setting $t_{j^*}$ to $1$ or $\sum_{i=0}^{N-1}s_i)$ is not significant in our experiments using the typical format of softmax output which is accurate to 8 decimal places.}, \textit{there is only one feasible optimal transport plan.}

In order to satisfy the criteria of ${\rm\textbf{W}}$, all masses should be transferred to the cluster of the ground truth label $j^*$, as shown in Fig. \ref{fig:r3}. Then, the Wasserstein distance between softmax prediction {\rm{\textbf{s}}} and one-hot target {\rm{\textbf{t}}} can be simplified to\begin{equation}
\mathcal{L}_{{\rm\textbf{D}}_{i,j}^{f}}({\rm{\textbf{s},\textbf{t}}})=\sum_{i=0}^{N-1} s_i f(d_{i,j^*}) \label{con:df}
\end{equation} where ${\rm\textbf{D}}_{i,j}^f=f(d_{i,j})$. $f$ is an increasing function proper, $e.g., p^{th}$ power of $d_{i,j}$ and Huber function. The exact solution of Eq. \eqref{con:df} can be computed with a complexity of $\mathcal{O}(N)$. The ground metric term $f(d_{i,j^*})$ works as the weights $w.r.t.$ $s_i$, which takes all classes into account following a soft attention scheme \cite{liu2018dependency,liu2019dependency,liu2019permutation}. It explicitly encourages the probabilities distributing on the neighboring classes of $j^*$. Since each $s_i$ is a function of the network parameters, differentiating $\mathcal{L}_{{\rm\textbf{D}}_{i,j}^{f}} w.r.t.$ network parameters yields $\sum_{i=0}^{N-1}s_i'f(d_{i,j^*})$.

In contrast, the cross-entropy loss in one-hot setting can be formulated as $-1{\rm log}s_{j^*}$, which only considers a single class prediction as hard attention \cite{liu2018dependency,liu2019dependency}. Similarly, the regression loss using softmax prediction could be $f(d_{i^*,j^*})$, where $i^*$ is the class with maximum prediction probability.

%% file: 4_Approach2.tex
\begin{figure}[t!]
\centering
\begin{tabular}{cc}
\includegraphics[height=4.8cm]{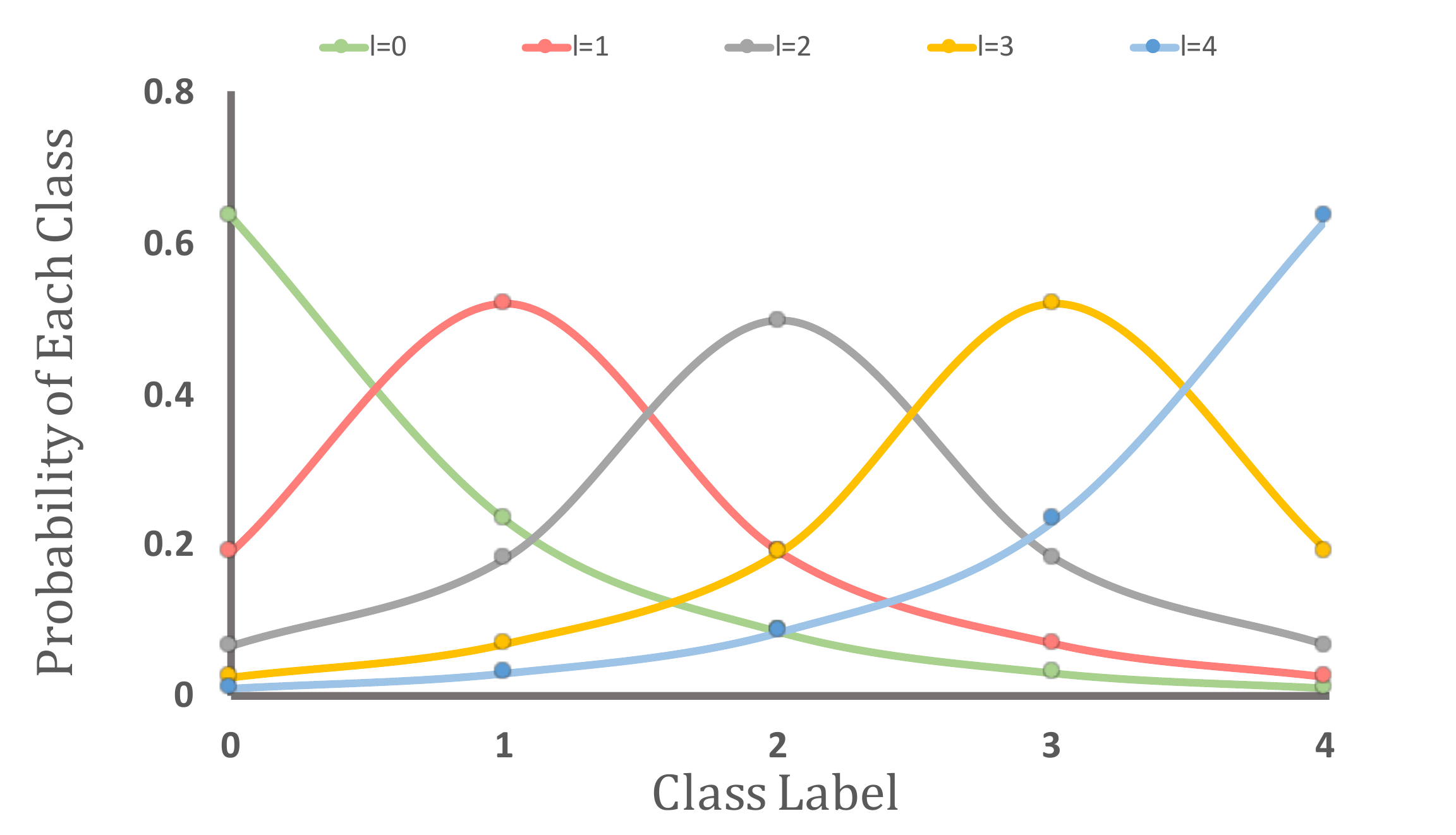}
\end{tabular}
\caption{The distribution of normalized exponential function $e^{-|i-l|}$ for a dataset with 5 classes.}
\label{fig:r4}
\end{figure}

\subsection{Unimodal-uniform label smoothing}

The outlier noise exists commonly in data-driven tasks, and can be modeled by a uniform distribution \cite{szegedy2016rethinking,liu2017adaptive,liu2018adaptive}. However, ordinal labels are more possible to be mislabeled as a close class of the true class. It is more reasonable to form a unimodal distribution to depict the inlier noise in health risk level estimation, which has a peak at class $j^*$ while decreasing its value for farther classes. This paper has sampled a continuous unimodal distribution and follow by normalization.

In here, we propose to sample on an exponential function $e^{\frac{-|i-l|}{\tau}}$ and processed by a softmax normalization. Discrete distributions with five classes are illustrated in Fig. \ref{fig:r4}. 

\begin{figure}[t]
\centering
\begin{tabular}{cc}
\includegraphics[height=4.9cm]{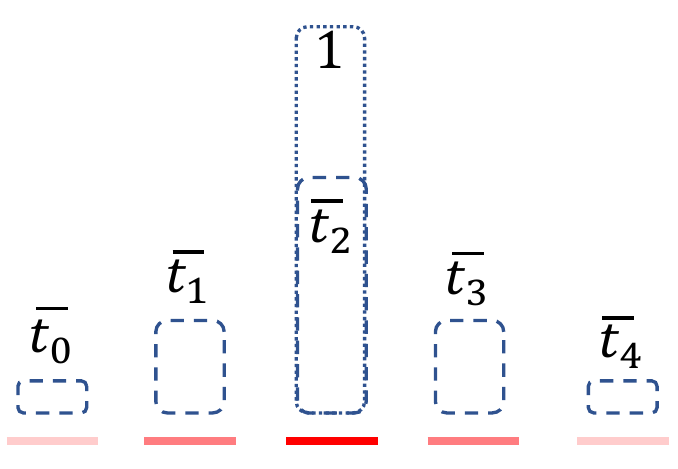}
\end{tabular}
\caption{The unimodal-uniform smoothed target label distribution when the ground-of-truth class is 2.}
\label{fig:r5}
\end{figure}

The normalized unimodal value is denoted as $p_j$. The unimodal-uniform smoothed target distribution ${\rm{{\overline{\textbf{t}}}}}$ is constructed by replacing $t_{j}$ in ${\rm\textbf{t}}$ with $(1-\xi-\eta)t_{j}+\xi p_j +\eta \frac{1}{N}$, which can be regarded as the weighted sum of the original label distribution ${\rm\textbf{t}}$ and a unimodal-uniform mixture distribution. In the context that the uniform distribution target is utilized for the CE loss, it is equivalent to label smoothing \cite{szegedy2016rethinking}, a typical mechanism for outlier noisy label training, which encourages the model to accommodate less-confident labels. The smoothed distribution is shown in Fig. \ref{fig:r5}

Noticing that the smoothed target label can also be adapted into CE loss which is formulated as

\begin{equation}\mathcal{L}=\sum_i^{N-1} \overline{{t}}_i [\texttt{-log}(p(y=i|\textbf{\texttt{x}}))]\end{equation}

By enforcing {\rm\textbf{s}} to form a unimodal-uniform mixture distribution, we also implicitly encourage the probabilities to distribute on the neighbor classes of $j^*$.

Since $\overline{{t}}_i$ is monotonically decreasing $w.r.t.$ the farther distance from the true class $l$, we can regard it as a weight of $\texttt{-log}(p(y=i|\textbf{\texttt{x}}))$. Since the target label regularization can be processed in advance, the training time does not increase by adding the unimodal-uniform mixture distribution regularization.

%% file: 4_Approach3.tex
\subsection{Wasserstein training with smoothed target}

When using the unimodal-uniform smoothed label, the fast solution in Eq. \eqref{con:df} does not apply. The possible solution is to regard it as a general case and solve its exact solution in the complexity higher than $\mathcal{O}(N^3)$ or approximate its exact solution with a complexity in $\mathcal{O}(N^2)$. In this section, a series of analytic formulation when the ground metric is a nonnegative increasing linear/convex/concave function $w.r.t.$ Euclidean distance with reasonable complexity.\\

\noindent\textbf{3.3.1 Linear and Convex function $ w.r.t.$ $d_{i,j}$ as the ground metric.}

Choosing the ground metric as $d_{i,j}$ or extend it as an nonnegative increasing and convex function of $d_{i,j}$, the analytic formulation of our loss function $\mathcal{L}_{d_{i,j}}(\rm{\textbf{s},\overline{\textbf{t}}})$ can be formulated as

\begin{equation}
\mathcal{L}_{d_{i,j}}{(\rm{{\textbf{s},\overline{\textbf{t}}}})}=\sum_{j=0}^{N-1}|{\sum_{i=0}^{j}(s_i-\overline{{t}}_i)}|\label{con:medi}
\end{equation}

Eq. \eqref{con:medi} was developed in \cite{werman1986bipartite}, where it is proved for sets of points with unitary masses on a line. A similar conclusion for the Kantorovich-Rubinstein problem was derived in \cite{cabrelli1995kantorovich,cabrelli1998linear}, which is known to be identical to the Wasserstein distance problem when ${{\rm\textbf{D}}_{i,j}}$ is a distance. We note that this is true for $\mathcal{L}_{d_{i,j}}$ (but not hold for $\mathcal{L}_{{\rm\textbf{D}}^{\rho}}{(\rm{{\textbf{s},\overline{\textbf{t}}}})}$ with $\rho>1$). An equivalent calculation is proposed from the cumulative distribution perspective \cite{rabin2009statistical}. All of these works notice that computing Eq. \eqref{con:medi} can be solved in linear time ($\mathcal{O}(N)$). See \cite{villani2003topics} for a comprehensive review).

Noticing that the partial derivative of Eq. \eqref{con:medi} $w.r.t.$ $s_n$ is $\sum_{j=0}^{N-1}{\rm{sgn}}(\varphi_j)\sum_{i=0}^{j}(\delta_{i,n}-s_i),$ where $\varphi_j=\sum_{i=0}^{j}(s_i-{\overline{t}_i}),$ and $\delta_{i,n}=1$ when $i=n$.

Here, we give some measures \footnote{We refer to ``measure'', since a $\rho^{th}$-root normalization is required to get a distance \cite{villani2003topics}, which satisfies positive definiteness, symmetry and triangle inequality.} using the typical convex ground metric function.

Using $d^\rho$ as ground metric, i.e., ${\rm\textbf{D}}_{i,j}^\rho= d_{i,j}^\rho$. The loss function is written as $\mathcal{L}_{{\rm\textbf{D}}_{i,j}^\rho}{(\rm{{\textbf{s},\overline{\textbf{t}}}})}$, with $\rho=2,3,\cdots$. When set $\rho=2$, the Wasserstein distance is equivalent to the Cram\'{e}r distance \cite{rizzo2016energy}. Note that the Cram\'{e}r distance is not a distance metric proper. However, its square root is.

We can also use Huber cost function with a parameter $\tau$ and denote it as $\mathcal{L}_{{\rm\textbf{D}}_{i,j}^{H\tau}}{(\rm{{\textbf{s},\overline{\textbf{t}}}})}$.\begin{equation}
{\rm\textbf{D}}_{i,j}^{H\tau}=\left\{
             \begin{array}{ll}
             d_{i,j}^2&{\rm{if}}~d_{i,j}\leq\tau\\
             \tau(2d_{i,j}-\tau)&{\rm{otherwise}}.\\
             \end{array}
             \right.
\end{equation}

\begin{figure*}[t!]
\centering
\includegraphics[height=4.3cm]{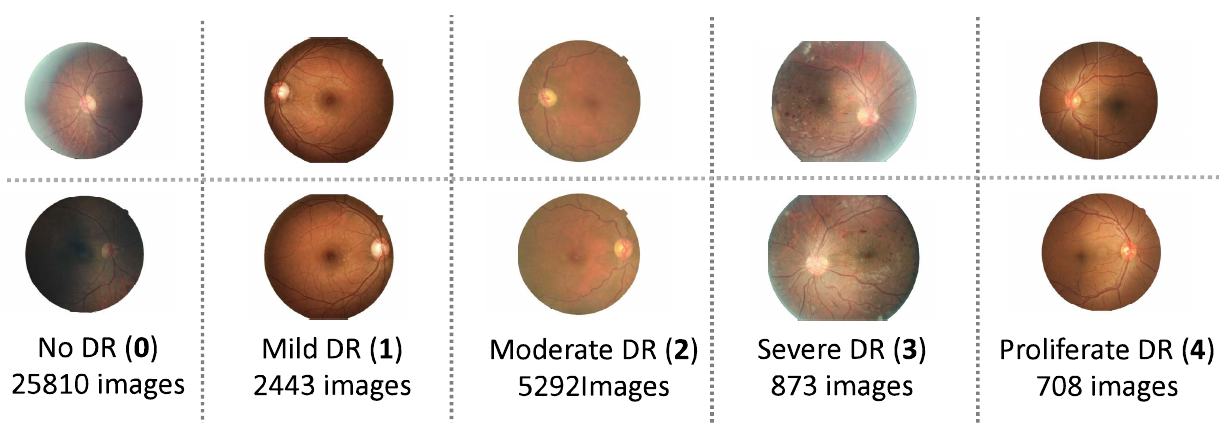}
\caption{\normalsize{Some samples with different retinopathy level in the DR dataset. The top row is the left retinopathy image while the bottom row is the right retinopathy image. The sampls show a large inner-class variation and small inter-class variation.}}
\label{fig:r7}
\end{figure*}

\noindent\textbf{3.3.3 Concave function $w.r.t.$ $d_{i,j}$ as the ground metric}

In practice, it may be useful to define the ground metric as a nonnegative, concave and increasing function $w.r.t.$ $d_{i,j}$. Although the general computation speed of the concave function is not satisfactory, the step function $f(t)=\mathbbm{1}_{t\neq 0}$ (one every where except at 0) can be a special case, which has significantly less complexity \cite{villani2003topics}. Assuming that the $f(t)=\mathbbm{1}_{t\neq 0}$, the Wasserstein metric between two normalized discrete histograms on $N$ bins can be simplified to the $\ell_1$ distance. \begin{equation}
\mathcal{L}_{\mathbbm{1}{d_{i,j}\neq 0}}{(\rm{{\textbf{s},\overline{\textbf{t}}}})}=\frac{1}{2}\sum_{i=0}^{N-1}{|{\rm{s}}_i-{\rm{\overline{t}}}_i|}=\frac{1}{2}||{\rm{\textbf{s}}}-{\rm{\overline{\textbf{t}}}}||_1
\end{equation}where $||\cdot||_1$ is the discrete $\ell_1$ norm.  

However, its fast computation is at the cost of losing the ability to discriminate the difference of probability in a different position of bins.

%% file: 5_Experiments.tex
\section{Experiments}

To evaluate the effectiveness of our Wasserstein loss, we show implementation details and experimental results on the two widely used health risk level diagnosis datasets, $i.e.,$ Diabetic Retinopathy and Ultrasound BIRADS datasets. To manifest the effectiveness of each setting choice and their combinations, we give a serial of elaborate ablation studies along with the standard measures. For the fair comparison, we choose the same neural network backbones as in previous works. All of networks in our training use the $\mathcal{L}_2$ norm of $10^{-4}$, ADAM optimizer \cite{kingma2014adam} with 128 training batch-size and initial learning rate of $10^{-3}$. The learning rate will be divided by ten when either the validation loss or the valid set QWK plateaus. There is no significant difference in the training time of Wasserstein loss and CE-loss based multi-class classification, and the smoothed unimodal target label is constructed before the training stage. All the experiments are implemented in deep learning platform Pytorch \footnote{\url{https://pytorch.org/}}. 

We use the prefix $\approx$ denote the approximate computation of Wasserstein distance \cite{cuturi2013sinkhorn,frogner2015learning}. ${(\rm{{\textbf{s},{\textbf{t}}}})}$ and ${(\rm{{\textbf{s},\overline{\textbf{t}}}})}$ refer to using one-hot or smoothed target label. For instance, $\mathcal{L}_{d_{i,j}}(\rm{\textbf{s},{\textbf{t}}})$ means choosing Wasserstein loss with Euclidean distance in a line as ground metric and using one-hot target label.

\subsection{Evaluations}

Since the health risk level has a discrete label, the performance of a system can be simply measured by the average classification accuracy as the conventional classification problem. \cite{ratner2018learning} further utilized the Mean True Negative Rate (TNR) at True Positive Rate (TPR) of 0.95. The relatively high TPR used here is fitted for strict TPR requirements of medical applications to avoid misdiagnosing diseased cases as healthy. However, they do not consider the severity of different misclassification.

Considering the inherent ordered label relationship, the Mean Absolute Error (MAE) metric, $i.e.,$ $L_1$ loss, can also be used as an evaluation metric in related risk evaluation datasets \cite{niu2016ordinal}, which is computed using the average of the absolute errors between the ground truth and the estimated result. Here, we also propose its use in evaluating the proposed method on two medical health risk evaluation benchmarks.

\begin{table*}[t]
\caption{\normalsize{Performance on the DR dataset.}}
\vspace{+10pt}
\label{tab:1}
\begin{center}
\begin{tabular}{|c|c|c|c|c|c|c|}
    \hline
   \multirow{2}*{Evaluations}& \multicolumn{3}{c|}{Mean TNR@TPR=0.95} & \multirow{2}*{Valid Acc} & \multirow{2}*{Valid QWK}& \multirow{2}*{MAE}\\ \cline{2-4}
    & 0 \textit{vs} 1-4 & 0-1 \textit{vs} 2-4 & 0-2 \textit{vs} 3-4 & & & \\ \hline \hline
      
    MC& 41.5\% &30.9\%& 31.1\% &82.4\%& 0.724&0.37 \\ \hline 
    RG& 40.3\% &30.6\%& 30.8 \%&76.2\%& 0.705&0.38 \\ \hline 
    Poisson \cite{beckham2017unimodal}& 38.8\% &30.0\%& 29.6 \%&77.1\%& 0.713&0.38 \\ \hline 

    MT \cite{ratner2018learning}& 42.7\% &31.7\%& 31.3\% &82.8\%& 0.726&0.36 \\ \hline
    SB\cite{liu2018ordinal}& 44.0\% &33.1\%& 32.6\% &84.2\%& 0.743&0.32 \\ \hline \hline

    $\mathcal{L}_{d_{i,j}}(\rm{\textbf{s},{\textbf{t}}})$& {46.9}\% &{37.1}\%& {34.4}\% &{87.3}\%& {0.768}&{0.29}\\\hline
    
    $\mathcal{L}_{{\rm\textbf{D}}_{i,j}^2}{(\rm{{\textbf{s},{\textbf{t}}}})}$& {47.2}\% &{37.3}\%& {34.6}\% &{87.4}\%& {0.769}&{0.28}\\\hline
    
    $\mathcal{L}_{{\rm\textbf{D}}_{i,j}^{H\tau}}{(\rm{{\textbf{s},{\textbf{t}}}})}$& {47.2}\% &{37.4}\%& {34.5}\% &{87.6}\%& {0.769}&{0.28}\\\hline   \hline

    MC${(\rm{{\textbf{s},\overline{\textbf{t}}}})}$& {42.4}\% &{31.2}\%& {31.8}\% &{82.7}\%& {0.728}&{0.35}\\\hline\hline
    
    $\approx\mathcal{L}_{d_{i,j}}(\rm{\textbf{s},\overline{\textbf{t}})}$& {45.8}\% &{36.4}\%& {33.8}\% &{86.6}\%& {0.759}&{0.29}\\\hline
    
    $\approx\mathcal{L}_{{\rm\textbf{D}}_{i,j}^2}{(\rm{{\textbf{s},\overline{\textbf{t}}}})}$& {45.8}\% &{36.5}\%& {33.9}\% &{86.5}\%& {0.760}&{0.29}\\\hline
    
    $\approx\mathcal{L}_{{\rm\textbf{D}}_{i,j}^{H\tau}}{(\rm{{\textbf{s},\overline{\textbf{t}}}})}$& {45.9}\% &{36.6}\%& {34.0}\% &{86.6}\%& {0.760}&{0.28}\\  \hline\hline  
    
    $\mathcal{L}_{d_{i,j}}(\rm{\textbf{s},\overline{{\textbf{t}}}})$& {47.3}\% &{37.5}\%& {34.8}\% &{87.8}\%& {0.771}&{0.27}\\\hline
    
    $\mathcal{L}_{{\rm\textbf{D}}_{i,j}^2}{(\rm{{\textbf{s},\overline{\textbf{t}}}})}$&\textbf{47.6}\% &\textbf{37.7}\%& \textbf{34.9}\% &{87.8}\%& {0.772}&\textbf{0.26}\\\hline
    
    $\mathcal{L}_{{\rm\textbf{D}}_{i,j}^{H\tau}}{(\rm{{\textbf{s},\overline{\textbf{t}}}})}$& {47.5}\% &\textbf{37.7}\%& {34.8}\% &\textbf{88.0}\%& \textbf{0.773}&\textbf{0.26}\\  \hline

\end{tabular}
\end{center}
\end{table*}

Moreover, as defined in previous Kaggle DR competition, we also evaluate the quadratic weighted kappa (QWK) \footnote{\url{https://www.kaggle.com/c/diabetic-retinopathy-detection/overview/evaluation}}. It can punish the misclassification proportional to the distance between the predicted label of the network and the ground-of-truth label \cite{cohen1968weighted}. The QWK is formulated as: \begin{align}
k=1-\frac{\sum_{i,j}{\textbf{W}_{i,j}{\textbf{O}_{i,j}}}}{\sum_{i,j}{\textbf{W}_{i,j}{\textbf{E}_{i,j}}}}
\end{align}

We evaluate on Diabetic Retinopathy (DR) and Ultrasound BIRADS datasets which are suitable for deep learning implementations in the medical area.

\subsection{Diabetic Retinopathy (DR)}

The Diabetic Retinopathy (DR) dataset \footnote{\url{https://www.kaggle.com/c/diabetic-retinopathy-detection}} contains a large amount of high-resolution fundus ($i.e.,$ interior surface at the back of the eye) images which have been labeled as five levels of DR. The level 0 to 4 representing the No DR, Mild DR, Moderate DR, Severe DR, and Proliferative DR, respectively. The left and right fundus images from 17563 patients are publicly available. The ResNet \cite{he2016deep} style model with 11 ResBlocks as in \cite{beckham2017unimodal,liu2018ordinal} has been adopted for DR dataset. We use five neurons with softmax normalization as our output to represent the probability of each level.

In our experiments, we follow the setting of \cite{beckham2017unimodal,liu2018ordinal}. The subject-independent 10-fold cross-validation is adopted, $i.e.,$ the validation set consisting of 10\% of the patients is set aside. The images belonging to a subject will only appear in a single fold. By doing this, we can avoid contamination. The images are also preprocessed as in \cite{beckham2017unimodal,liu2018ordinal,liu2019unimodala} and subsequently resized as $256\times 256$ size images. Some examples can be found in Fig. \ref{fig:r7}.

We show the results in the DR dataset in Table \ref{tab:1}. The evaluation metrics discussed earlier is utilized.  Several baseline methods are chosen for comparisons. For example, the CE-loss based multi-class classification (MC), MSE-loss based metric regression (MSE), Poisson distribution output with CE-loss (Poisson), multi-task network using a series of binary CE loss (MT), and the stick-breaking with CE-loss (SB).

\begin{figure*}[t!]
\centering
\includegraphics[height=4cm]{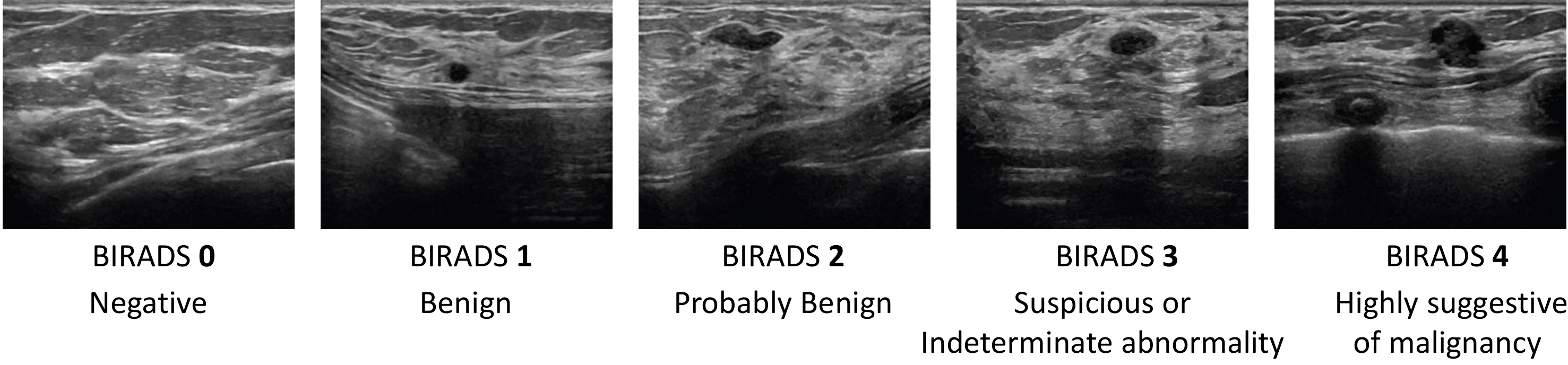}
\caption{{Some samples with different malignant risk in the US-BIRADS.}}
\label{fig:r8}
\end{figure*}

\begin{table*}[t]
\caption{{Performance on the US-BIRADS dataset.*Our implementations have slightly higher TNR using MC baseline than the results reported in \cite{ratner2018learning}}}. 
\label{tab:2}
\begin{center}
\begin{tabular}{|c|c|c|c|c|c|c|}
    \hline
   \multirow{2}*{Evaluations}& \multicolumn{3}{c|}{Mean TNR@TPR=0.95} & \multirow{2}*{Valid Acc} & \multirow{2}*{Valid QWK}& \multirow{2}*{MAE}\\ \cline{2-4}
    & 0 \textit{vs} 1-4 & 0-1 \textit{vs} 2-4 & 0-2 \textit{vs} 3-4 & & &\\ \hline \hline
      
    MC& 33.2\%* & 28.7\%*& 29.8\%* &73.3\%& 0.678&0.42\\ \hline 
    RG& 31.6\%&28.5\%& 29.5\% &73.0\%& 0.677& 0.44\\ \hline 
    Poisson \cite{beckham2017unimodal}& 29.6\% &27.2\%& 29.5\% &72.2\%& 0.665&0.45 \\ \hline 
    MT \cite{ratner2018learning}& 38.5\% &29.2\%& 31.3\% &76.5\%& 0.685&0.41 \\ \hline
    SB\cite{liu2018ordinal}& 39.1\% &30.2\%& 32.0\% &78.3\%& 0.694&0.39 \\ \hline \hline
    
    $\mathcal{L}_{d_{i,j}}(\rm{\textbf{s},{\textbf{t}}})$& {42.5}\% &{33.6}\%& {35.7}\% &{80.1}\%& {0.712}&{0.36}\\\hline
    
    $\mathcal{L}_{{\rm\textbf{D}}_{i,j}^2}{(\rm{{\textbf{s},{\textbf{t}}}})}$& {42.6}\% &{33.8}\%& {35.9}\% &{80.2}\%& {0.714}&{0.35}\\\hline
    
    $\mathcal{L}_{{\rm\textbf{D}}_{i,j}^{H\tau}}{(\rm{{\textbf{s},{\textbf{t}}}})}$& {42.6}\% &{33.7}\%& {35.9}\% &{80.3}\%& {0.715}&{0.35}\\\hline\hline

    MC${(\rm{{\textbf{s},\overline{\textbf{t}}}})}$& {33.4}\% &{29.0}\%& {30.4}\% &{73.6}\%& {0.682}&{0.40}\\\hline\hline
    
    $\mathcal{L}_{d_{i,j}}(\rm{\textbf{s},\overline{{\textbf{t}}}})$& {42.9}\% &{34.0}\%& {36.2}\% &{80.5}\%& {0.715}&{0.34}\\\hline
    
    $\mathcal{L}_{{\rm\textbf{D}}_{i,j}^2}{(\rm{{\textbf{s},\overline{\textbf{t}}}})}$&\textbf {43.0}\% &\textbf{34.2}\%&\textbf {36.3}\% &{80.5}\%&\textbf {0.716}&{0.34}\\\hline
    
    $\mathcal{L}_{{\rm\textbf{D}}_{i,j}^{H\tau}}{(\rm{{\textbf{s},\overline{\textbf{t}}}})}$& \textbf{43.0}\% &{34.1}\%& \textbf{36.3}\% &\textbf{80.6}\%&\textbf {0.716}&\textbf{0.33}\\  \hline

\end{tabular}
\end{center}
\end{table*}

The MC usually outperforms MSE in most of the metric. However, MSE usually appears to be competitive w.r.t. MAE, since MSE optimizes a similar metric as MAE in its training phase. The Poisson does not manage to achieve performance improvements in most of the evaluations due to its uncontrollable variance. The MT is more promising than MC as it considers the successive relationship, despite it has a lot to be tuned hyper-parameters. By addressing some limitation in MT, the SB has a better performance than MT. 

Our Wasserstein training outperforms all of the previous methods, especially the $\mathcal{L}_{{\rm\textbf{D}}_{i,j}^2}$ and $\mathcal{L}_{{\rm\textbf{D}}_{i,j}^{H\tau}}$ which use the convex function of the Euclidean distance in a line as the ground metric.

Moreover, the unimodal-uniform smoothed target label can efficiently improve the performance without additional training costs. The smoothing process is benefiting to both conventional CE loss and Wasserstein loss. Besides, the exact solution can outperform its approximate counterpart consistently.

We set our hyper-parameters  $\xi=0.15$, $\eta=0.05$ and $\tau=1$. QWK is not sensitive to the $\tau \in \left\{0.8,0.9,1,1.1\right\}$ when we fix the $\xi=0.15$. Similarly, the QWK keep at the same level when we adjust $\xi$ from 0.12 to 0.18.

\subsection{Ultrasound BIRADS}

The second medical dataset is the Ultrasound BIRADS (US-BIRADS) \cite{ratner2018learning}. It consists of 4904 breast images with the BIRADS system label. Considering the relatively limited number of samples in level 4, we usually regard the 3-4 as a single level \cite{ratner2018learning}. That results in 2700 healthy (0) images, 1113 benign (1) images, 359 probably benign (2) images, and 732 may contain/contain malignant (3-4) images. We divide this dataset into 5 subsets for subject-independent five-fold cross-validation. We show some samples at different levels in Fig. \ref{fig:r8}.

AlexNet style architecture \cite{krizhevsky2012imagenet} with six convolution layers and following two dense layers is used for US-BIRADS image dataset as in \cite{ratner2018learning}. We set $\xi=0.15$, $\eta=0.05$, and $\tau=1$ for the unimodal distribution.

The leading performance of our method is also observed in the US-BIRADS dataset (Table \ref{tab:2}). Since its labels are noisier (more severe annotator-dependent problem), the unimodal-uniform smoothing usually offers a more appealing contribution to the results. The Wasserstein training with convex ground metric function and smoothed target label achieves state-of-the-art performance consistently.

%% file: 6_Conclusions.tex
\section{Conclusions}
Based on the Wasserstein distances, we proposed an efficient loss function as an alternative to the CE loss for health risk level estimation. The ground metric inherits the inter-class correlation and can be predefined by with an increasing function w.r.t. the Euclidean distance of a line. The intlier and outlier noise in health risk data can be incorporated in a unimodal-uniform mixture distribution to form the smoothed target. We systematically discusses the fast closed-form solutions in one-hot and conservative label cases. The results show that the best performance can be achieved by choosing convex function, unimodal-uniform distribution for smoothing and solving its exact solution. Although it was originally developed for health risk level estimation, it is essentially applicable to other problems with discrete and ordinal labels. In the future, we intend to develop an adaptive ground metric learning scheme, and adjust the shape of conservative target distribution automatically.

\section{Acknowledgement}
The funding support from National Institute of Health (NIH), National Institute of Neurological Disorders and Stroke (NINDS) (NS061841, NS095986), Youth Innovation Promotion Association, CAS (2017264), Innovative Foundation of
CIOMP, CAS (Y586320150) and Hong Kong Government General Research Fund GRF (Ref. No.152202/14E) are greatly appreciated.